\documentclass{ws-ijbc}
\usepackage{ws-rotating}     
\usepackage{graphicx}
\usepackage{epstopdf}
\usepackage{epsfig}
\usepackage{amsmath}
\usepackage{amsfonts}
\usepackage{amssymb}
\usepackage{subfigure}
\usepackage{psfrag}
\usepackage{color}
\usepackage[english]{babel}
\usepackage[english]{varioref}
\usepackage{soul}
\begin{document}
\catchline{}{}{}{}{} 
\markboth{D.W.C. Marcondes, G.F. Comassetto, B.G. Pedro,
  J.C.C. Vieira, A. Hoff, F. Prebianca, C. Manchein and
  H.A. Albuquerque}{Extensive numerical study and circuitry
  implementation of the Watt governor model} 
\title{Extensive numerical study and circuitry implementation of the
  Watt governor model}  
\author{D.W.C. Marcondes, G.F. Comassetto, B.G. Pedro, J.C.C. Vieira, 
    A. Hoff, F. Prebianca, \\ C. Manchein$^{\dagger}$, and
    H.A. Albuquerque$^{\ast}$} 
\address{Physics Department, Universidade Do Estado de Santa Catarina, 
  Joinville, SC, 89219-710, Brazil\\
${}^{\ast}$holokx.albuquerque@udesc.br\\
${}^{\dagger}$cesar.manchein@udesc.br}
\maketitle
\begin{history}
\received{(to be inserted by publisher)}
\end{history}
\begin{abstract}
    In this work we {carry} out extensive numerical study of a 
    Watt-centrifugal-governor system model, and we also 
    {implement} an electronic circuit by analog computation to
    experimentally solve the model. Our numerical results show the
    existence of self-organized {\it stable periodic structures}  
    (SPSs) on parameter-space of the largest Lyapunov exponent and
    isospikes of time series of the Watt governor system model. 
    A peculiar hierarchical organization and period-adding bifurcation
    cascade of the SPSs are observed, and this self-organized cascade
    accumulates on a periodic boundary. {It is also} shown that
    the periods of these structures organize themselves obeying the
    {\it solutions} of Diophantine equations. In addition, an
    experimental setup is implemented by a circuitry analogy of 
    mechanical systems using analog computing technique to
    characterize the robustness of our numerical results. After
    {applying} an active control of chaos in the experiment, the
    effect of intrinsic experimental noise was minimized such that,
    the experimental results are in astonishing well agreement with
    our numerical findings. We can also mention as another remarkable
    result, the application of analog computing technique to perform
    an experimental circuitry analysis in real mechanical problems.
\end{abstract}



\section{Introduction}
\label{intro}
\noindent

\begin{figure}[htb]
    \centering
    \includegraphics*[width=7.0cm,angle=0]{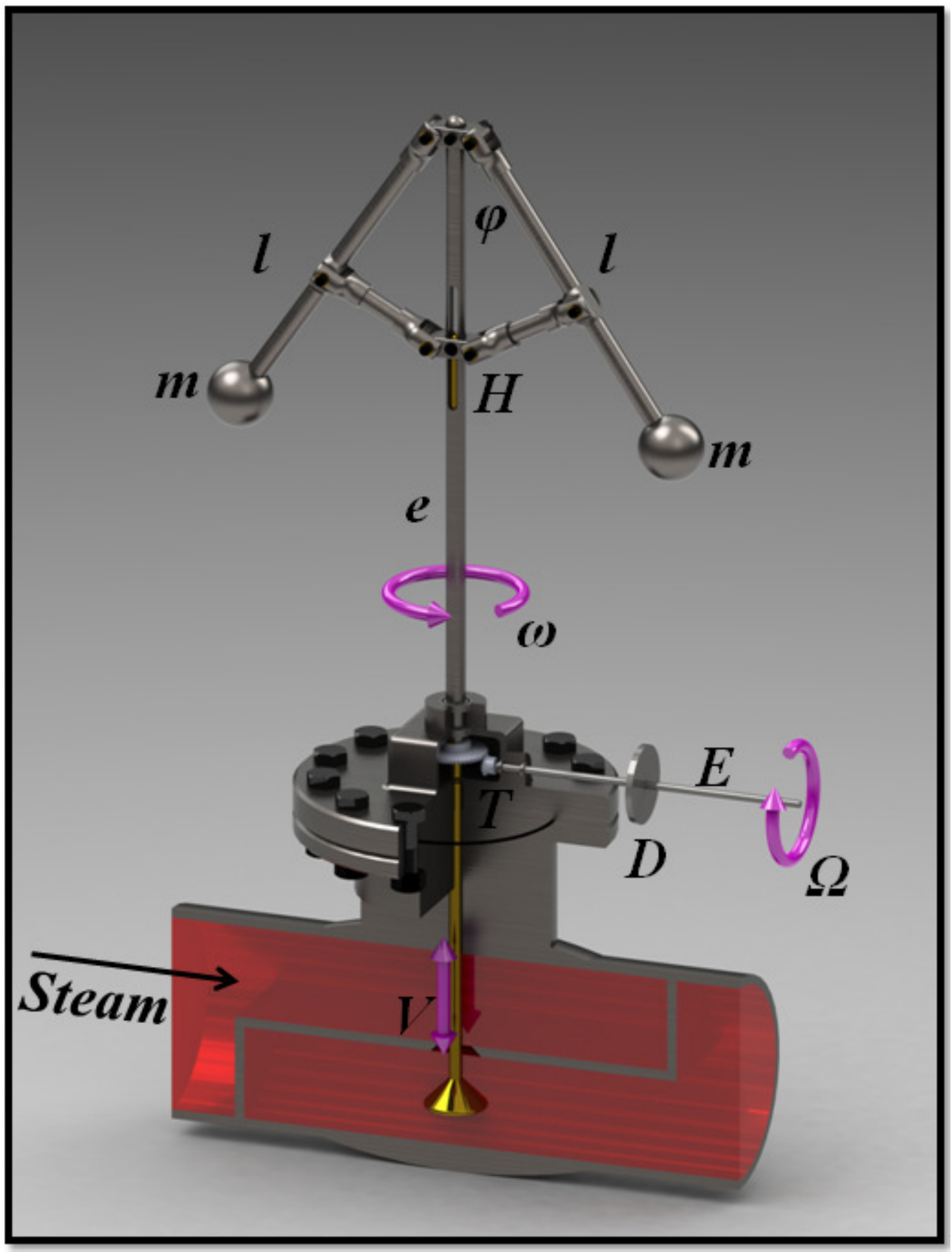}
    \caption{Schematic illustration of mechanical Watt governor 
      system coupled to a steam pipe.} 
    \label{fig:fig1}
  \end{figure}

The steam engine {played} an important role in the Industrial
Revolution \cite{h89} (see also references therein) at the end of the
$18^{th}$ century in Great Britain and it can be considered the
starting point for the automatic control theory \cite{m79}. In
a tentative to automatically control the steam flux into the engine
from the boiler, James Watt invented a centrifugal governor system
\cite{m68}, here called by  {\it Watt governor system} (WGS). 
Figure~\ref{fig:fig1} illustrates the mechanical WGS connected to a
valve that regulates the steam flux to the engine. In this case, as
the angular velocity of the Watt governor increases, the kinetic
energy of the balls increases. This allows the two masses on lever
arms {move} vertically outwards and upwards. If this motion goes far
enough, a security mechanism is initialized and it goes to reduce the
rate of working-fluid entering to the engine reducing its velocity,
preventing the over-velocity. 

The WGS is often used in the nonlinear sciences as an example of a
{dynamical} system. Indeed, due {to} the rich variety of complex  
{behaviors} and historical importance the dynamics of WGS is
studied here through of two different ways, as follows: (i) a
numerical analysis, by solving the equations of motion with both 
Runge-Kutta and numerical continuation methods and, (ii) an
experimental study, using {a circuitry} setup. These two ways to 
study {the} dynamics of WGS are self-complementary as shown {at
  next}. In the first one, a characterization of dynamics of WGS by
numerical simulations can be performed by {integration of} the
equations of motion while, in the second an analog computation
technique can be used to construct a circuit setup which works, based
on the equations of motion, in order to reproduce the dynamics of the
original mechanical system. A real experiment with a mechanical setup
using sensors coupled to WGS can also be achieved, although, it is not
the goal {here}. However, the control parameters in the set
of differential equations used to model the dynamics of WGS can also be
seen as sensors for measurements of dynamical parameters in the real
mechanical experiment. {Before applying} a detailed
discussion on numerical and experimental results, it is important
to introduce an appropriated theoretical model for WGS, used in our 
investigations. 

Applying the {Newton's} Second Law to translational and rotational
motion of the mechanical Watt governor (shown in Fig.~\ref{fig:fig1}),
we obtain the following set of coupled differential {equations}
\begin{eqnarray}
  \left\{
    \begin{array}{ll}
      \dot \varphi = \dfrac{d\varphi}{d\tau} = \psi, \\
      \\
      \dot \psi = \dfrac{d\psi}{d\tau} = c^{2}\Omega^{2}\sin \varphi
      \cos\varphi - \dfrac{g}{l}\sin \varphi -
      \dfrac{b}{m}\psi, \label{watt1}
      \\
      \\ 
      \dot \Omega = \dfrac{d\Omega}{d\tau} = \dfrac{1}{I}(\mu \cos \varphi
      - F), 
    \end{array}
  \right.
\end{eqnarray}
where $\varphi \in \left(0,\frac{\pi}{2} \right)$ is the angle of
deviation of the arms of the governor from its vertical axis $e$,
$\Omega \in [0,\infty)$ is the angular velocity of the rotation of the
flywheel $D$, $\omega$ is the angular velocity of $e$, $l$ is the
length of the arms, $m$ is the mass of each ball, $H$ is a sleeve
which supports the arms and slides along $e$, $T$ is a set of
transmission gears, $V$ is the valve that determines the supply of
steam to the engine, $\tau$ is the time, $\psi=d\varphi / d\tau$, $g$
is the standard acceleration of gravity. The angular velocity of
$e$ is related to the angular velocity of the rotation of the
flywheel $D$ by $\omega = c \Omega$, where $c$ is the positive
constant transmission ratio. $b$ is a positive constant of the
frictional force of the system, $I$ is the moment of inertia of the
flywheel, $F$ is an equivalent torque  of the load and $\mu$ is a
positive proportionality constant. The parameters $b,c$ and
$\mu$ only assume positive values to preserve the physical meaning
of Eqs.~(\ref{watt1}). For a derivation of Eqs.~(\ref{watt1}), the
reader is referred to Pontryagin~\cite{p62}.      

Based on Ref.~\cite{smb07}, we rewrite the set of Eqs.~(\ref{watt1})
in normalized and dimensionless form as follows     
\begin{eqnarray}
  \left\{
    \begin{array}{ll}
      \dot x = \dfrac{dx}{dt} = y, \\
      \\
      \dot y = \dfrac{dy}{dt} = z^{2}\sin x \cos x - \sin x -
      \varepsilon y, \\
      \\
      \dot z = \dfrac{dz}{dt} = \alpha(\cos x - \beta),
      \label{watt2} 
    \end{array}
  \right.
\end{eqnarray}
where
\begin{equation}
x = \varphi, \qquad y = \sqrt{\frac{l}{g}}\psi, \qquad z =
\sqrt{\frac{l}{g}}\Omega, \qquad \tau = \sqrt{\frac{l}{g}}t, \notag
\end{equation}
and
\begin{equation}
\varepsilon = \frac{b}{m}\sqrt{\frac{l}{g}}, \qquad \alpha =
\frac{cl\mu}{gI}, \qquad \beta = \frac{F}{\mu}, \notag
\end{equation}
for $\varepsilon > 0$, $\alpha > 0$ and, $0 < \beta < 1$. As
presented in the next section, those three parameters are varied in
pairs to perform a characterization of dynamics of WGS. For example,
when $\epsilon$ and $\alpha$ are varied, keeping $\beta$ fixed, it
means that, in a possible mechanical experiment of Fig.~\ref{fig:fig1}
(not realized in present work), the length $l$ of the arms and the
mass $m$ of each ball {has} been varied simultaneously. {As a
  consequence,} the parameters $\epsilon$, $\alpha$ and $\beta$ only
assume positive values to preserve {their} physical
{meanings}. The same analysis can be extended to other parameter
pairs combination (as shown bellow). {Numerical} and experimental
result sections {show that}, even  tiny changes in one or two of
those parameter values {are} enough to drastically change the
dynamics of mechanical WGS, {\it i.e.}, the analysis about how {their}
dynamical behaviors changes when those three parameters are varied in
pairs endorses the importance of the present study. Therefore, the
system~(\ref{watt2}) is a three-parameter nonlinear dynamical system
and extensive analyzes of stability and Hopf bifurcation in that
system {have been} reported in Ref.~\cite{smb07} (see also
references therein). {Analytical} and numerical {studies} in a
hexagonal governor system {can also be found} in
Ref.~\cite{zmcla10}.    

Once that system~(\ref{watt2}) is an autonomous three-dimensional
continuous system, chaotic dynamics can arise \cite{s01} and a
recent interest is on the description of its dynamics in
parameter-planes {simultaneously} varying two control parameters
\cite{cmab11,cmab14,g15,mepzgg16,srsab16,chgml16,g16}. To construct
those parameter-planes, it is usually used the Largest Lyapunov
exponent (Lyapunov diagram)
\cite{g15,mepzgg16,srsab16,chgml16,g16,r16} or the number of spikes in
one period of the orbits (isospikes diagram) \cite{g16,hsma13} (both
measures are properly defined at next). Those measures are identified
by color intensities on parameter-planes. Regardless
of the used measures, the two-parameter behavior of a large class of
nonlinear systems usually presents a complex bifurcation scenario and
generic {\it Stable Periodic Structures} (SPSs) \cite{cmab14} (see
also references therein), namely  {\it shrimps-shaped domains},
embedded in chaotic regions on the parameter-space. In some systems
these structures organize themselves with specific bifurcation
cascades, for example, period-adding cascades, and along preferred
directions on the parameter-space \cite{g15,srsab16,g16}.    

In this work, we report the existence of such SPSs organizing
themselves {along specific} directions on the Lyapunov and the
isospike diagrams of the WGS model~(\ref{watt2})~\cite{smb07}, 
using two invariant measures of the system namely, the {\it Largest
  Lyapunov Exponent} (LLE) and the {\it number of spikes} $q$ in one
period of the time-series of the attractors, respectively. {The
  results} presented here, corroborate the generic nature of those
SPSs, independently of the systems' nonlinearity. For example, these
SPSs are observed in piecewise-linear~\cite{srsab16,hsma14},
polynomial~\cite{cmab14,svors15,g16}, and exponential~\cite{car09}
nonlinearities. Here, both quadratic and trigonometric
{nonlinearity are} considered, since the model studied is derived
from a mechanical (steam) engine through the Newton's Second Law of
motion. Another numerical result presented in this work concerns to the
self-organization of the number of spikes $q$ of the SPSs, which we
can associate with solutions of Diophantine equations 
{that} are polynomial equations that {allow} the variables to be
integers only \cite{s91}. Bifurcation curves, using numerical
continuation {method} (used in {\tt MatCont} package)
\cite{dgk03,bbss11,bbds11}, are reported in the model with the curves
overlapped on the Lyapunov diagrams as previously performed for other
dynamical systems studies \cite{hsma13,hsma14}. Furthermore, an
experimental study of the system~(\ref{watt2}), implemented by a
{\it circuitry analogy} of mechanical systems using analog computing
technique, is also presented. The effects of noise in the experiment
are discussed with active control of chaos corroborating quite well
{with numerical results}. 

This work is organized as follows. In Section~\ref{numres}, we present  
the Lyapunov and the isospike diagrams, besides bifurcation
curves of system~(\ref{watt2}), obtained by numerical methods with
some discussions. Section~\ref{expres} is devoted to introduce the
circuitry implementation of the model and discuss the experimental
results besides {their} comparison with numerical
findings. Main concluding remarks are given in Section~\ref{rem}.   

\section{Numerical results}
\label{numres}
\begin{figure*}[htb]
  \centering
  \includegraphics*[width=0.95\linewidth,angle=0]{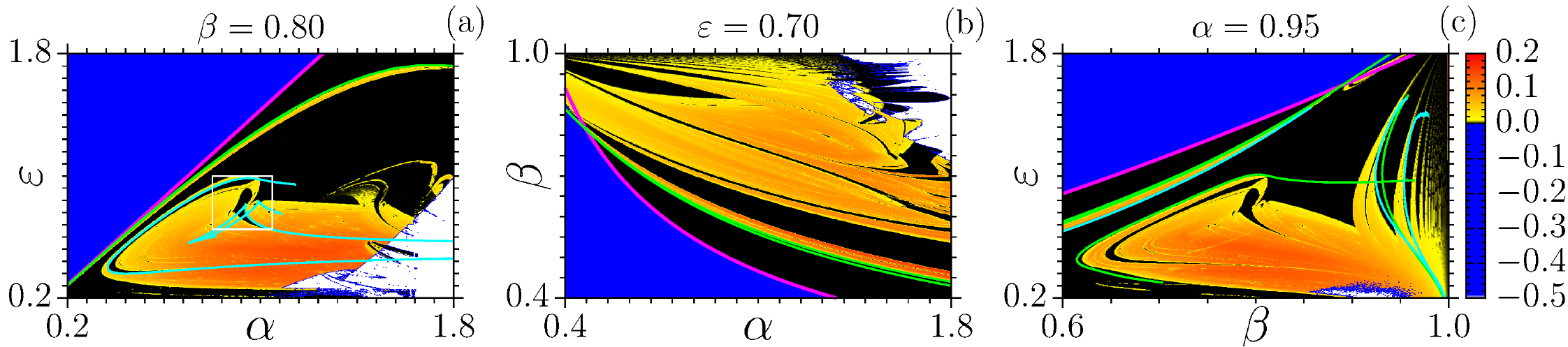}
  \caption{\protect Global view of Lyapunov diagrams (parameter
    planes) of system~(\ref{watt2}) for (a) $(\alpha,\varepsilon)$,
    (b) $(\alpha,\beta)$, and (c) $(\beta,\varepsilon)$. Black color
    indicates null LLEs or periodic behavior, yellow to red {one
      indicates} positive LLEs or chaotic behavior (see the color bar
    in (c)). The white region indicates divergence while the blue one
    equilibrium points. Bifurcation curves are overlapped on the
    diagrams to characterize the positions where there exist specific
    bifurcations. The white box in (a) delimits the amplification
    region plotted in Fig.~\ref{fig:fig3}(a).}   
  \label{fig:fig2}
\end{figure*}

Numerically solving the system~(\ref{watt2}) with the fourth-order
Runge-Kutta method, fixed time step-size equal {to} $10^{-3}$ and
iteration time of $10^6$, we evaluate the {\it Lyapunov
  spectra} and plot the largest Lyapunov exponent (LLE) for each point
of a grid of $600 \times 600$ values of three distinct parameter pairs:
$(\alpha, \beta)$, $(\alpha,\varepsilon)$, and
$(\beta,\varepsilon)$. Then, for each parameter-pair, we plot $3.6
\times 10^5$ LLEs to construct the Lyapunov diagrams using the
parameter grid in the axes and assigning colors to the
LLEs. Figure~\ref{fig:fig2} shows the Lyapunov diagrams for the
following parameter-planes: (a) $(\alpha,\varepsilon)$ with $\beta =
0.80$, (b) $(\alpha,\beta)$ with $\varepsilon = 0.70$, and (c) the
$(\beta,\varepsilon)$ with $\alpha = 0.95$. The parameter ranges
{are} chosen respecting the interval values used in Section~\ref{intro} at
the same time optimizing the visualization of different types of
behaviors for the WGS.  An intensity color scale of the LLEs is
defined as black, for null exponents (related to periodic dynamics
  of WGS), varying continuously to yellow up to red representing the 
positive exponents (chaotic dynamics), as plotted in the color 
bar in  Fig.~\ref{fig:fig2}(c). White color identifies set of
parameters where the system~(\ref{watt2}) diverges, and blue
represents the equilibrium point regions. The bifurcation
curves are overlapped on the Lyapunov diagrams and they are obtained
by numerical continuation \cite{hsma14,dgk03}. Specifically, magenta
curves are Hopf bifurcations, green ones are period-doubling
bifurcations and cyan are saddle-node bifurcations.   

  \begin{figure*}[htb]
    \centering
    \includegraphics*[width=0.95\linewidth,angle=0]{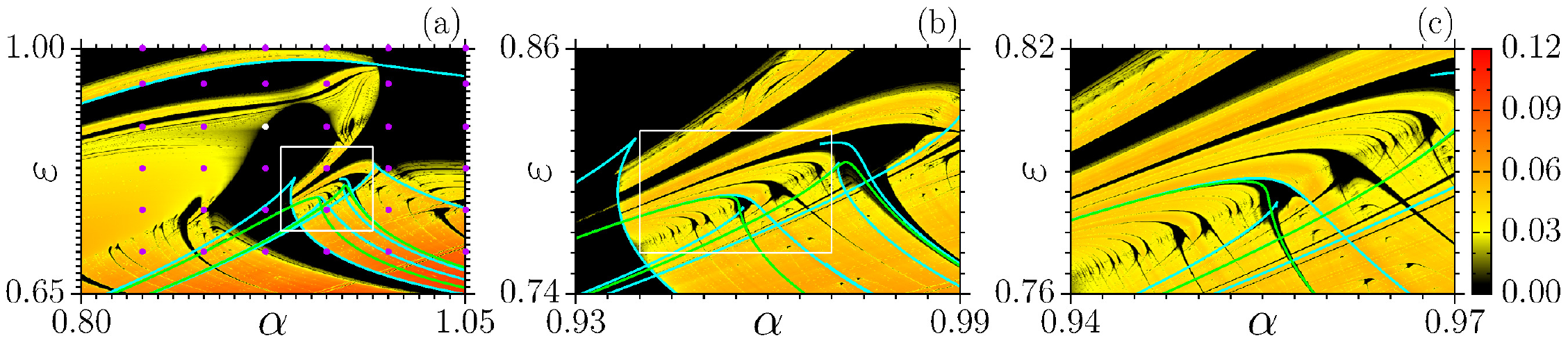}
    \caption{(a) Magnification of the white box in
      Fig.~\ref{fig:fig2}(a). Each magenta bullet plotted in this
      figure represents the parameter-pair used to {obtain} the
      experimental attractor projection (see Fig. ~\ref{fig:fig8}) of
      system~(\ref{watt2}). {(b) Magnification of the white box in 
        Fig.~\ref{fig:fig3}(a) and, (c) white box in
        Fig.~\ref{fig:fig3}(b), respectively}. The bifurcation curves
      are also overlapped on the diagrams. The left black region in (b),
      delimited by the cyan curve (saddle-node bifurcation) is an {\it
        accumulation horizon} where the stable periodic structures
      (better visualized in (c)), at right, accumulate themselves. The
      color scheme is shown in (c).}  
    \label{fig:fig3}
  \end{figure*}

  \begin{figure*}[htb]
    \centering
    \includegraphics*[width=0.95\linewidth,angle=0]{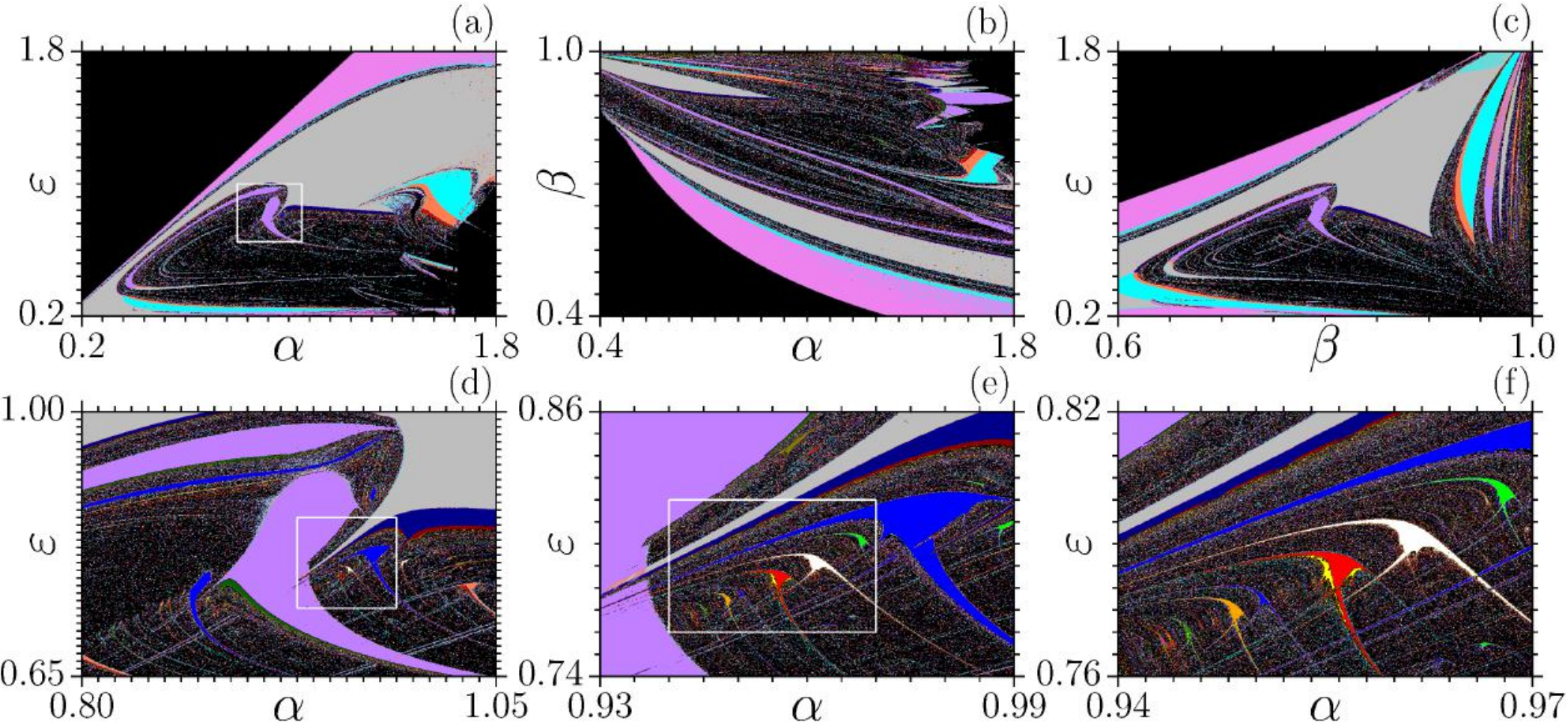}
    \caption{Isospike diagrams of
      system~(\ref{watt2}). Top row: (a)-(c) isospike diagrams of
      Fig.~\ref{fig:fig2}. Bottom row: (d)-(f) isospike diagrams of
      Fig.~\ref{fig:fig3}. The color scheme used is the following:
      violet for $q = 1$, cyan for $q = 2$, gray for $q = 3$, coral
      for $q = 4$, purple for $q = 5$, dark-blue for $q = 6$, red for
      $q = 8$, dark-green for $q = 10$, blue for $q = 11$, dark-red
      for $q = 12$, forest-green for $q = 13$, yellow for $q = 16$,
      white for $q = 17$, orange for $q = 21$, green for $q = 23$,
      brown for $q = 27$, magenta for $q = 31$, and navy for $q =
      41$. Black represents other values of $q$, equilibrium points,
      chaos and divergences.} 
    \label{fig:fig4}
  \end{figure*}

  \begin{figure*}[htb]
    \centering
    \includegraphics*[width=0.95\linewidth,angle=0]{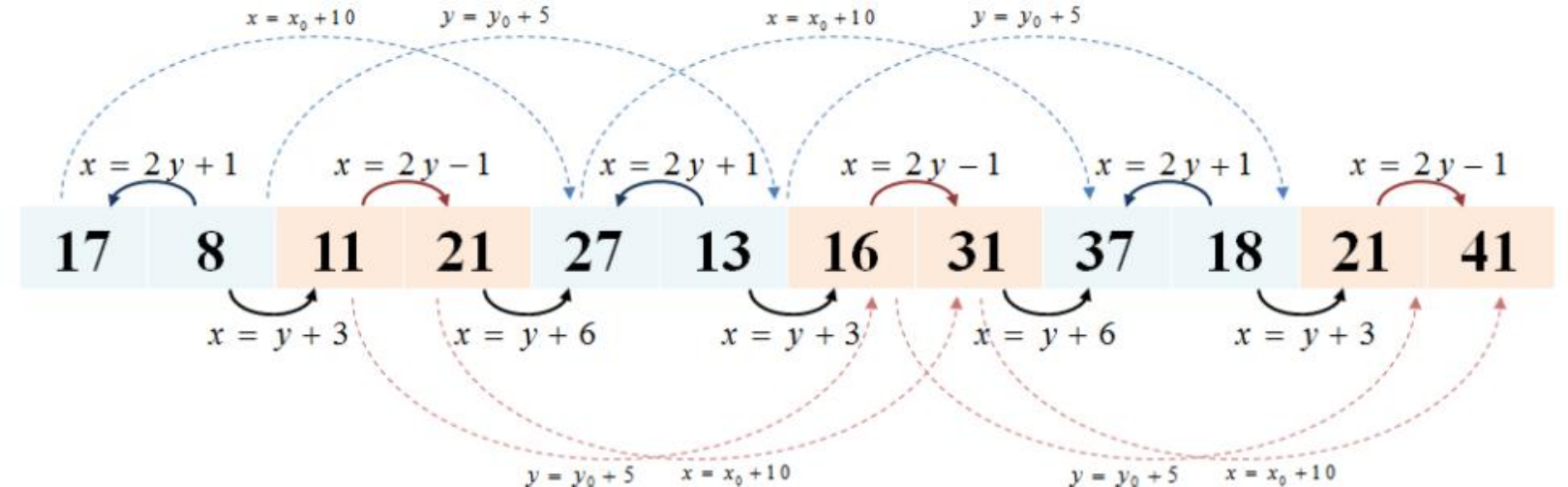}
    \caption{Schematic diagram showing the sequence of
      spikes observed in Fig.~\ref{fig:fig4}(f) and the organization
      rules of the structures obeying linear Diophantine equations,
      solid arrows (see the text), and spike-adding bifurcation
      cascades, dashed arrows. $x$ is the higher and $y$ is the lower
      number of spikes, $q$, of the pairs.}
    \label{fig:fig5}
  \end{figure*}

All the three projections $(\alpha,\beta)$, $(\alpha,\varepsilon)$,
and $(\beta,\varepsilon)$, present some basic features, as portions of
large periodic regions including SPSs embedded in chaotic
domains. Large regions of equilibrium points are also observed 
as well as divergence regions. The bifurcation curves present a quite
good agreement with the LLEs plotted in the Lyapunov diagrams and reveal
the type of bifurcations that occur in the system for specific
parameter-planes. In this case, we also can emphasize that such
information about bifurcation neighborhoods is not accessed only
by the LLEs. For example, the bifurcations between equilibrium 
points (blue domains) and periodic attractors (black domains) occur by 
Hopf bifurcations (see magenta curves) \cite{hsma14,dgk03} while the
period-doubling bifurcations between two periodic attractors and,
bifurcations between periodic attractors and chaotic ones occur by
saddle-node bifurcations, illustrated by green and cyan curves,
respectively.   

The existence of SPSs embedded in chaotic domains 
{is better visualized} in the diagrams plotted in Fig.~\ref{fig:fig3}, that are 
magnifications of portions of Lyapunov diagrams of the white box in 
Fig.~\ref{fig:fig2}(a) {and,} show the presence of shrimp-shaped
domains (stable-periodic-structures) organized themselves in some
specific directions in Lyapunov diagrams as {can be seen} in
Fig.~\ref{fig:fig3}(a). In Fig.~\ref{fig:fig3}(b), we show the
magnification of the white box in (a), and in Fig.~\ref{fig:fig3}(c)
the magnification of the white box in (b). The bifurcation curves are
also showed in those diagrams, with green and cyan curves being
period-doubling and saddle-node bifurcations, respectively. Regarding
those set of bifurcation curves, it is worth to note four of them. The
first is the cyan curve at the lowest right portion of the diagram in
(a). This curve borders the right portion of a SPS of {\it hook}
shape. In (b) we show that the curve delimits the {\it accumulation
  horizon} \cite{bg08}, where the SPSs organizing themselves towards
that periodic boundary. It is clear that in the {\it accumulation
  horizon} the bifurcations between the periodic and chaotic domains
are saddle-nodes. The others three curves are shown in
Fig.~\ref{fig:fig3}(c). They are the three curves overlapped the {
  shrimp-shaped-domain} in the center of the diagram. Those curves are
two {saddle-node} and one period-doubling. In Ref.~\cite{hsma13} the
authors showed that the endoskeleton of a { shrimp-shaped-domain}
{is} formed by four bifurcation curves, two saddle-node, and two
period-doubling. Here, in (c), it was impossible to obtain the second
period-doubling curve due to the numerical precision in the numerical
continuation method \cite{dgk03,bbss11,bbds11}. But, it is clear the
good agreement between the numerical integration (Lyapunov diagram)
and continuation (bifurcation curves) methods.  
  \begin{figure*}[htb]
    \centering
    \includegraphics*[width=0.95\linewidth,angle=0]{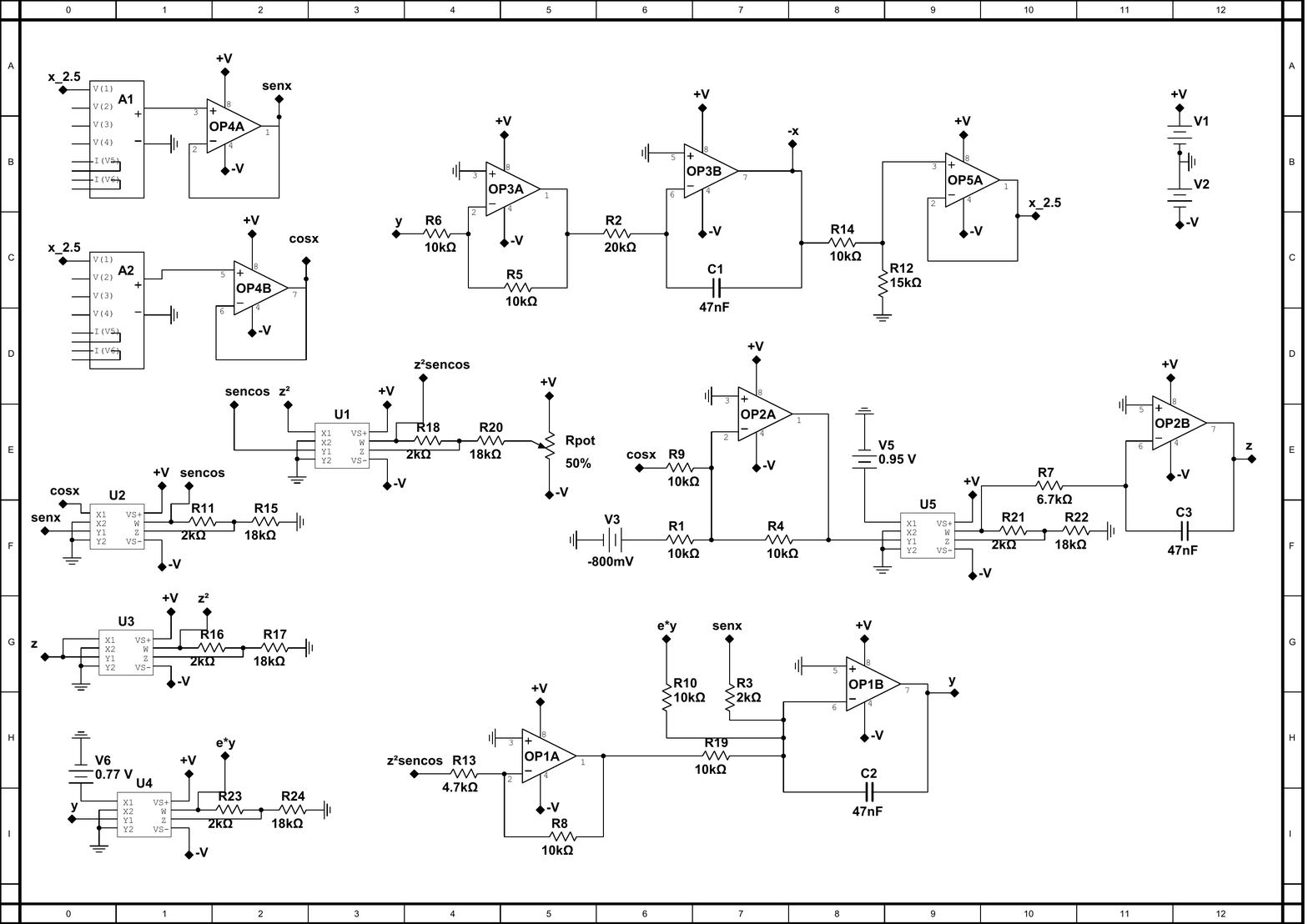}
    \caption{Schematic circuitry implementation of the Watt governor
      model based on the set of Eqs.~(\ref{watt2}).}
    \label{fig:fig6}
  \end{figure*}

Figure~\ref{fig:fig3}(c) was plotted to study in more details the
self-organization of the SPSs described in the paragraph above. 
{Such SPSs} organize themselves towards the
periodic boundary, with the border delimited by a saddle-node
bifurcation curve. Regarding this self-organization, {we can}
associate to it the pattern of pairs of structures, and these pairs 
also organize themselves towards the periodic boundary. At this point,
it is important to mention that a common self-organization of SPSs was
previously reported in a large class of dynamical systems  
\cite{cmab14,g15,chgml16,g16,hsma13,hsma14,car09,bg08}. Specifically
considering the sequence plotted in Fig.~\ref{fig:fig3}(c) we see that
it is a period-adding sequence, where the SPSs organize themselves
such that the number of spikes ($q$), in one period of the time series
of the attractors related with these structures, increases by a
constant factor along a specific direction. Furthermore, in the case
reported in Fig.~\ref{fig:fig3}(c), we observe a peculiar sequence
of accumulation where the number of spikes $q$ emerges organized
according to a solution of integers of {\it Diophantine
  equations}. This is better explained in the following.    

In Fig.~\ref{fig:fig4} we present the isospike diagrams of
system~(\ref{watt2}) in meshes of $1200 \times 1200$ parameter 
values. Those diagrams are parameter-planes in which the number $q$ of
spikes, in one period of time of the time-series, is used instead of
LLEs, as in Figs.~\ref{fig:fig2} and \ref{fig:fig3}. The first row,
Figs.~\ref{fig:fig4}(a)-(c), shows the isospike diagrams related to
Fig.~\ref{fig:fig2}, and below, Figs.~\ref{fig:fig4}(d)-(f), the
isospike diagrams of Fig.~\ref{fig:fig3}. {The peculiar} sequence
of accumulation mentioned above, can be explained via those isospike
diagrams. In Figs.~\ref{fig:fig4}(e) and (f), the pairs of SPSs are
better visualized, for example, the white-red pair of structures with
$q=(17;8)$, and the blue-orange pair with $q=(11;21)$,
and so on, until to the accumulation boundary structure with $q=5$,
the purple SPS at left in  Fig.~\ref{fig:fig4}(e). Since the $q$
number of those structures {is} known, one can now analyze their
organization rules. To make easier our analysis, Fig.~\ref{fig:fig5}
shows a schematic diagram with the pairs sorted in an inverse sequence
as plotted in Fig.~\ref{fig:fig4}(f),  
where each pair represents a pair of SPSs in this same figure. The
{curved} arrows with {equations} represent the organization rules of the
SPSs' spikes. Based on this diagram we suppose that there are at least
four distinct rules for the organization of the SPSs, namely two
rules for organization of the pairs, upper solid arrows in
Fig.~\ref{fig:fig5}, and two rules connecting the distinct pairs,
lower solid arrows in Fig.~\ref{fig:fig5}. The {equations} that
represent the organization rules can be viewed as {\it linear
  Diophantine equations}. However, the two {equations} for
organization of the spikes of distinct pairs can be described with
only one {equation}, {\it i.e.}, with the quadratic Diophantine
equation $(x-2y)^2=1$, where $x$ and $y$ are the higher- and
lower-spikes of the pairs, respectively. The two {equations} that
connect the distinct pairs can also be described with only one
quadratic Diophantine equation $(2x-2y-9)^2=9$. Therefore,
with the two quadratic Diophantine equations {it is} possible to
construct the whole accumulation sequence of SPSs towards the
accumulation boundary. In this diagram, the spike-adding bifurcation
cascades between alternated pairs become evident, {\it i.e.}, between  
alternated solutions of the quadratic Diophantine equation
$(x-2y)^2=1$, as represented by the upper and the lower dashed
arrows. The {equations} for the spike-adding cascades indicate that we
have a {\it general} solution for the spikes $x$ and $y$ given
respectively by $x = x_0 + 2n$ and $y = y_0 + n$, where $n = 5$. To
summarize this discussion {it is} essential to emphasize that the 
accumulation boundary of this sequence has $q=5$, as shown by
Fig.~\ref{fig:fig4}(e). 

We finish this Section stating the following partial conclusions: the
dynamics of WGS, modeled by the set of Eqs.~(\ref{watt2}) in
normalized and  dimensionless variables, has a sensitive dependence of
the combination of parameter values of $\epsilon$, $\alpha$ and
$\beta$. As shown by our numerical results, depending on this
parameter-set the WGS presents periodic (with high-spike values) and
chaotic dynamics as shown respectively, by black and yellow up to red
domains in Figs.~\ref{fig:fig2} and \ref{fig:fig3}. In a physical
point of view, the mechanical WGS (shown in Fig.~\ref{fig:fig1})
presents periodic and  chaotic motions depending on the choice of
{the} three-parameter values. Besides, our results clearly {show}
the existence of self-organized SPSs on parameter-space of the largest
Lyapunov exponent and isospikes of time series of the WGS.  In
particular, a peculiar hierarchical organization and period-adding
bifurcation cascade of the SPSs are observed and, {they are}
accumulating on a periodic boundary (see Fig.~\ref{fig:fig4}). More
interestingly, it is shown that the periods of these structures
organize themselves obeying the solutions of Diophantine equations.  

\section{Experimental results}
\label{expres}
After an extensive numerical investigation about the dynamics of WGS,
we also perform a quite interesting experimental study of same
system, proposing an analog circuit implementation of
system~(\ref{watt2}), as shown in Fig.~\ref{fig:fig6}. We start this
Section describing the main devices used to construct the experimental
setup as presented at next.  

To construct the experimental setup it was used the following
devices: (i) the operational amplifiers $TL074$
($OP1A$,$\ldots$,$OP2A$,etc) {to} perform the addition,
subtraction, and integration operations; (ii) the analog 
multipliers $AD633$ ($U1$,$U2$,$\ldots$) {to implement} the
multiplications, and the trigonometric function converters $AD639$
($A1$ and $A2$), {to perform} the trigonometric operations. In this
circuitry implementation (of system~(\ref{watt2})), the control
parameters $\varepsilon$, $\alpha$ and, $\beta$ can be independently
varied by adjusting the bias $V5$, $V3$, and $V6$, respectively. We
have chosen vary bias \cite{rm09,mr14} instead of resistors
\cite{srsab16,tapbff16,vraor10,vraor12}, which are commonly used in
circuitry implementations of dynamical systems, for a better fine
tuning that we can obtain in the parameters. Besides, commercial and
digital potentiometers are useful in some applications in which the
signal-noise relation is not relevant, for instance, in linear
systems. On the other hand, for nonlinear systems in which chaotic
dynamics {are} significant, the signal-noise relation should be
considered. As far as we know, commercial or home-made (as the digital
\cite{srsab16}) potentiometers, are noise sources in circuitry
implementations, and in nonlinear dynamical systems, noise signal with
a large enough strength may perturb the periodic attractors
\cite{vraor10,vraor12}, leading the system to chaotic motion. As
presented bellow, the circuitry implementation of system~(\ref{watt2})
{is} highly noise-sensitive. Therefore, it is imperative to minimize
the effects caused by sources of noise because, in this case, the
system becomes experimentally more stable. 

The bias sources (or the control parameters $\alpha$, $\varepsilon$
and, $\beta$) are controlled by the data acquisition ($DAQ$) board
$NI$~$PCIe-6259$, by its analog outputs. In the measurements, we also
use the $DAQ$ board, using three analog inputs (the variables $x$,
$y$, and $z$). We used the {\tt Python} ambient to control the $DAQ$
board measurements and also to analyze the data from time series with
the {\tt TISEAN} package \cite{hks99}.  

\begin{figure}[htb]
    \centering
    \includegraphics*[width=0.60\linewidth,angle=0]{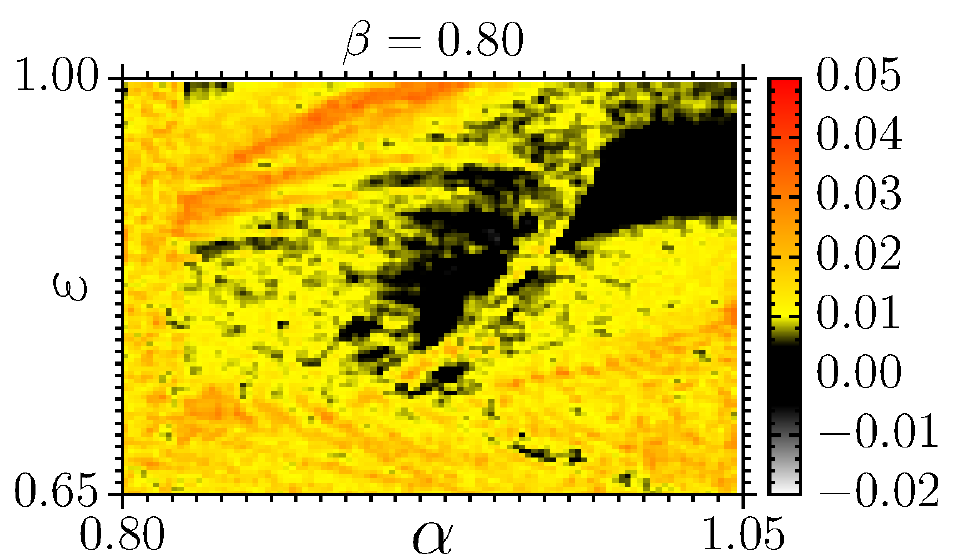}
    \caption{Lyapunov diagram of the circuitry
      implementation of the system~(\ref{watt2}). This diagram is the
      experimental version of the numerical diagram of
      Fig.~\ref{fig:fig3}(a). The color scheme for the LLEs is in the
      right-side palette.}  
    \label{fig:fig7}
  \end{figure}

After run a set of realizations, we plotted the experimental Lyapunov   
diagram in Fig.~\ref{fig:fig7}, obtained by the circuitry
implementation of the system~(\ref{watt2}), in a resolution of $10^2
\times 10^2$. The color scheme used is the same as discussed in
Fig.~\ref{fig:fig2}: Black and yellow up to red colors for periodic
and chaotic motions, respectively. This diagram is the experimental
version of the diagram shown in Fig.~\ref{fig:fig3}(a). To obtain this 
experimental diagram, the circuit of Fig.~\ref{fig:fig6} was built in
a single-sided printed circuit board ($PCB$) driven by a $DC$
symmetric power-source, and connected to the $DAQ$ board, in the
computer, by a block of $BNC$ connectors. {Suitable} home-made
routines in {\tt Python} ambient were developed to {automate} the
measurements of the variables $x$, $y$, and $z$, and after to analyze
the data using the routine  {\tt lyap\_spec} from {\tt TISEAN} package
\cite{hks99}.    

  \begin{figure*}[!htb]
    \centering
    \includegraphics*[width=0.95\linewidth,angle=0]{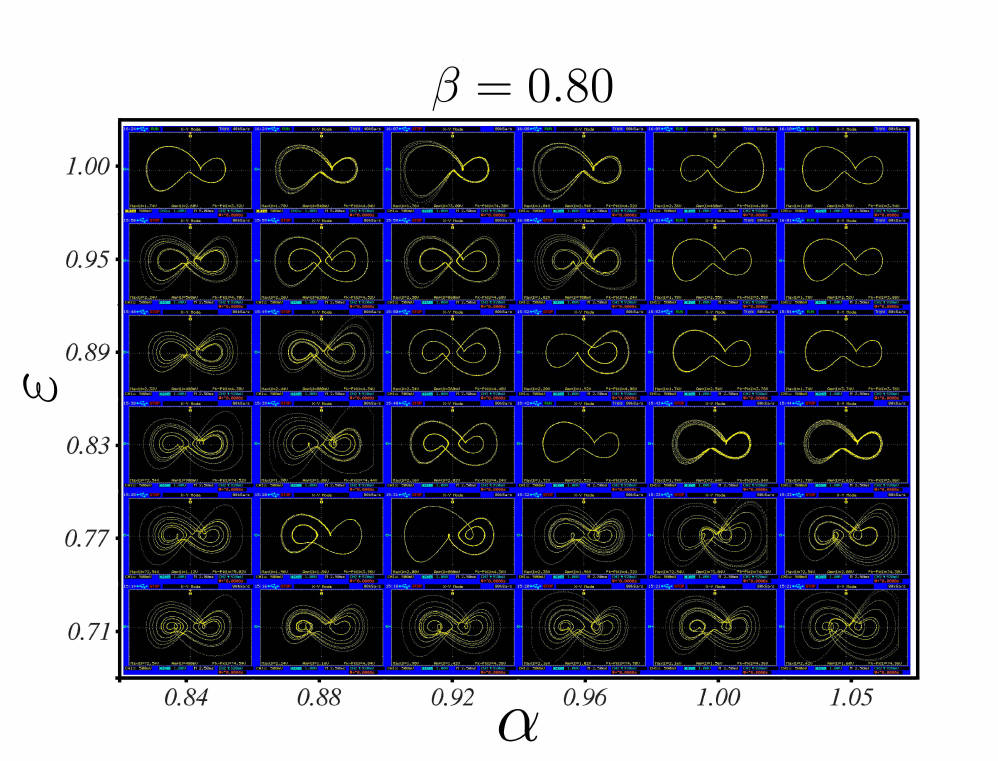}
    \caption{Each panel in this figure is an experimental
      attractor's projection in the plane of variables $ x\times y$
      for the parameter pair highlighted by a magenta bullet in
      Lyapunov diagram of Fig.~\ref{fig:fig3}(a). The periodic
      attractors were obtained applying chaos control procedure (see
      the text) in some points of the experimental Lyapunov diagram of
      Fig.~\ref{fig:fig7}.}    
    \label{fig:fig8}
  \end{figure*}

As shown and discussed elsewhere \cite{vraor10,vraor12,bglmm00}, noise 
{disturbs the basin of attractions of periodic attractors of}
nonlinear systems \cite{mcb13}, usually {inducing} periodic
behaviors to chaotic ones if the noise strength is 
high-enough, and specifically in the parameter-spaces, deforming and
destroying the periodic structures \cite{ham17}. That effect is shown in
Fig.~\ref{fig:fig7}  compared with the Fig.~\ref{fig:fig3}(a). We
claim that noise in the circuitry implementation of Fig.~\ref{fig:fig6} 
destroyed the smallest SPSs observed in Fig.~\ref{fig:fig3}(a) and
deformed the two biggest ones: The central {\it hook-shaped} structure
and the {top-right} periodic region, as shown in Fig.~\ref{fig:fig7}.    

To {reduce} the noise effects in the circuit, and to control some
chaotic domains that arise in the experimental Lyapunov diagram, we
apply an effective chaos control procedure, as proposed in
  Ref.~\cite{bglmm00}, in the circuitry implementation by adding a 
precision commercial potentiometer ($Rpot$ in Fig.~\ref{fig:fig6}) to
stabilize the unstable periodic orbits (UPOs) embedded in the stable chaotic
attractor for some set of parameters. With this procedure, we show in
Fig.~\ref{fig:fig8} an experimental diagram of some exemplary
experimental attractors presenting the dynamics associated to the
  parameter-pairs highlighted by magenta bullets in
  Fig.~\ref{fig:fig3}(a) {where this set of attractors is} obtained 
  by setting the circuit in certain values of parameters (each magenta
  bullet plotted in Fig.~\ref{fig:fig3}(a), {is} associated to an
  attractor in Fig.~~\ref{fig:fig8}). {To obtain such results it}
  was necessary a manually fine tuning of the potentiometer, to
  stabilize the UPOs,  as the attractors' data were recorded directly
  from the digital oscilloscope by using print screen
  functionality. With this control, we are able to stabilize the
  experimental UPOs with the same number of spikes $q$, as shown in 
Fig.~\ref{fig:fig4}(d), from the chaotic orbits born from noise
perturbations. For example, the periodic attractor for the parameter
pair (illustrated by white bullet in Fig.~\ref{fig:fig3}(a))
$(\alpha,\varepsilon)=(0.92,0.89)$ in Fig.~\ref{fig:fig8} has $q = 5$
{\it i.e.} the same $q$ for the purple {\it hook-shaped} structure
plotted in Fig.~\ref{fig:fig4}(d). Although the process, of
manually fine tuning the potentiometer, used to stabilize the
UPOs seems rough or imprecise, it was very useful as shown by the
remarkable concordance between results plotted in
Figs.~\ref{fig:fig3}(a) and \ref{fig:fig8}, {\it i.e} bullets
plotted on yellow up to red/black color regions in
Fig.~\ref{fig:fig3}(a) related to chaotic/periodic attractors in
Fig.~\ref{fig:fig8}. In addition, we also show in this exemplary set
of panels how the dynamics of the mechanical WGS ({in which} the
dynamics is reproduced by an electronic circuit) changes (from
periodic to chaotic or {\it vice-versa}) when at least one control
parameter is varied.   

The second set of partial conclusions is based on a rich and complex
dynamics presented by the quite interesting experimental study of WGS
through of analog circuit implementation of system \ref{watt2}, as
shown in Fig.~\ref{fig:fig6}. In this circuitry implementation, the
control parameters $\epsilon$, $\alpha$ and, $\beta$ can be
independently varied by adjusting bias in the circuit. After
a set of experimental realizations we plotted the experimental
Lyapunov diagram, which is in agreement with the theoretical
counterpart. We also applied an effective chaos control procedure, as
proposed in Ref.~\cite{bglmm00}, in the circuitry implementation by
adding a precision commercial potentiometer (Rpot in
Fig.~\ref{fig:fig6}) to stabilize the unstable periodic orbits (UPOs)
embedded in the stable chaotic attractor for some sets of parameters.

\section{Summary and conclusions}
\label{rem}
In this work we carried out extensive numerical and
experimental studies in a model of the {\it Watt governor system}
(WGS). We start to investigate the dynamics of WGS applying a
numerical approach to characterize the presence of generic 
{\it stable periodic structures} (SPSs) embedded in chaotic domains of
the three-dimensional parameter-space of system
\ref{watt2}. {Numerical results corroborate with the {\it generic}
  nature of those SPSs, independently of the systems' nonlinearity.}
Using the numerical continuation method to solve the same system we 
characterize the Hopf, period-doubling and saddle-node bifurcation
curves, as plotted in the parameter planes of Figs.~\ref{fig:fig2}
and \ref{fig:fig3}. Comparing the position of such curves in these 
{Lyapunov diagrams} with the results shown in the isospike diagrams
(see Fig.~\ref{fig:fig4}) {it is} possible to realize the remarkable
concordance between such results, even if the invariant measure used
in each case is different. We also discuss {how the SPSs} organizes
themselves in some preferred directions on the Lyapunov and  isospike
diagrams  of the WGS model (see Eqs.~(\ref{watt2})).  {We
  characterize a new hierarchical organization of these SPSs, where
  the periods follow the  solutions of {\it Diophantine
    equations}}. It is worth to mention that in other portions of the
Lyapunov diagrams we detected sequences of structures which periods
are self-organized following the solutions of Diophantine
equations. To the best of our knowledge, the numerical results
reported here, through the Lyapunov and spike diagrams, are difficult
(maybe impossible) to be analytically predicted. One {open}
question that {deserves} further studies is what {sorts} of
systems present the hierarchical organization shown here, {\it i.e.},
following the solutions of Diophantine equations. 

To test the robustness of our numerical results an experimental setup 
was implemented by a circuitry analogy of mechanical systems using an
analog computing technique. After {applying} an active control of
chaos in the experiment, the effect of intrinsic experimental noise
was minimized such that, the experimental results {are} in
astonishing well agreement with our numerical findings. Another
interesting result  obtained from our experimental approach is related
to the application of analog computing technique to mechanical
problems that can be used to {obtain} experimental results from
different real problems.

\nonumsection{Acknowledgments}
The authors thank Conselho Nacional de Desenvolvimento Cient\'\i
fico e Tecnol\'ogico (CNPq), Coordena\c c\~ao de Aperfei\c{c}oamento
de Pessoal de N\'\i vel Superior (CAPES), Funda\c c\~ao de Amparo \`a 
Pesquisa e Inova\c c\~ao do Estado de Santa Catarina (FAPESC),
Brazilian agencies, for financial support.




\end{document}